\documentclass[conference]{IEEEtran}
\IEEEoverridecommandlockouts
\usepackage{amsmath,amssymb,amsfonts}
\usepackage{algorithmic}
\usepackage{graphicx}
\usepackage{textcomp}
\usepackage[dvipsnames]{xcolor}
\usepackage[utf8]{inputenc}
\usepackage[perpage,symbol]{footmisc}
\usepackage[english]{babel}
\usepackage{multirow,multicol}
\usepackage{amsmath, bm}
\usepackage{pbox, booktabs, rotating}
\usepackage{subfigure}
\usepackage{array, makecell, lipsum}
\usepackage{enumerate}
\usepackage{threeparttable}
\usepackage{multirow,hyphenat,balance}
\usepackage{placeins}
\usepackage[compress]{cite}
\usepackage{float,color}
\usepackage{ragged2e}
\usepackage{enumerate}
\usepackage{etoolbox}
\usepackage{paralist}
\usepackage{cases}
\usepackage{url}
\usepackage{framed}
\usepackage{flushend}
\usepackage{amsmath}
\usepackage{amsthm}
\usepackage{amsthm}
\usepackage[english]{babel}
\usepackage{amssymb}
\usepackage[caption=false,font=normalsize,labelfont=sf,textfont=sf]{subfig}
\usepackage[linesnumbered,ruled]{algorithm2e}
\let\oldnl\nl% Store \nl in \oldnl
\newcommand{\nonl}{\renewcommand{\nl}{\let\nl\oldnl}}% Remove line number for one line
\usepackage{pifont}
\usepackage{commath}

\newcommand*{\eg}{\textit{e.g.,}\@\xspace}

\newcolumntype{M}[1]{>{\centering\arraybackslash}m{#1}}
\usepackage{tikz}
\newcommand*\circled[1]{\tikz[baseline=(char.base)]{
            \node[shape=circle,draw,inner sep=1.2pt] (char) {#1};}}

\usepackage[all=normal,paragraphs=tight,wordspacing=normal,charwidths=tight,indent=normal,floats=tight,leading=normal,lists=tight]{savetrees}

\def\BibTeX{{\rm B\kern-.05em{\sc i\kern-.025em b}\kern-.08em
    T\kern-.1667em\lower.7ex\hbox{E}\kern-.125emX}}
    
\title{Beware of Discarding Used SRAMs: Information \\is Stored Permanently}

\author{\IEEEauthorblockN{Joshua Hovanes, Yadi Zhong, and Ujjwal Guin}
\IEEEauthorblockA{Department of Electrical and Computer Engineering, Auburn University \\
Auburn, AL, USA \\
Emails: \{josh.hovanes, yadi, ujjwal.guin\}@auburn.edu}
}

\begin{document}

\thispagestyle{plain}
\pagestyle{plain}

\maketitle

\begin{abstract}
Data recovery has long been a focus of the electronics industry for decades by security experts, focusing on hard disk recovery, a type of non-volatile memory. Unfortunately, none of the existing research, neither from academia, industry, or government, have ever considered data recovery from volatile memories. The data is lost when it is powered off, by definition. To the best of our knowledge, we are the first to present an approach to recovering data from a static random access memory. It is conventional wisdom that SRAM loses its contents whenever it turns off, and it is not required to protect sensitive information, e.g., the firmware code, secret encryption keys, etc., when an SRAM-based computing system retires. Unfortunately, the recycling of integrated circuits poses a severe threat to the protection of intellectual properties. In this paper, we present a novel concept to retrieve SRAM data as the aging of SRAMs leads to a power-up state with an imprint of the stored values. We show that our proposed approaches can partially recover the previously used SRAM content. The accuracy of the recovered data can be further increased by incorporating multiple SRAM chips compared to a single one. It is impossible to retrieve the prior content of some stable SRAM cells, where aging shifts these cells towards stability. As the locations of these cells vary from chip to chip due to uncontrollable process variation, the same cell has a higher chance of being unstable or stable against aging in any of the chips, which helps us recover the content. Finally, majority voting is used to combine a set of SRAM chips' data to recover the stored data. We present our experimental result using commercial off-the-shelf SRAM chips with stored binary image data before performing accelerated aging. We demonstrate the successful partial retrieval on SRAM chips that are aged with as little as 4 hours of accelerated aging with 85$^{\circ}$C. 
\end{abstract}

\vspace{5px}
\begin{IEEEkeywords}
SRAM, volatile memories, aging, data retrieval, IP theft.
\end{IEEEkeywords}

\section{Introduction} \label{sec:intro}

Integrated circuits (ICs) are at the heart of virtually every industry, from commercial, industrial, to defense sectors. It is crucial for a nation to maintain technological superiority, especially in semiconductors, over any potential adversaries~\cite{DoD2015challenge, semicon_1, CHIPS, SIA2020, ezell2021moore, whitehouse2021, collier2021zero}. However, with the evolving globalization in the electronics supply chain, the fabrication, test, assembly, and packaging are performed offshore. Bureau of Industry and Security reports that over 43\% of electronic products' assembly is performed not by the original equipment manufacturer (OEM), but is outsourced to third-party contractors~\cite{BIS2022} due to the close proximity with IC manufacturers, lower labor cost, and potential subsidies~\cite{ilo2017}. The majority of electronics are now manufactured and assembled in environments with limited trust, government oversight, or visibility, posing a serious threat to protecting intellectual properties (IPs). IP theft, where an adversary obtains an IP illegally, has been the prime focus since globalization started, and researchers from academia, industry, and the government have proposed different solutions for IP protection.

\begin{figure}[t!]
\centering 
\includegraphics[width=1\linewidth]{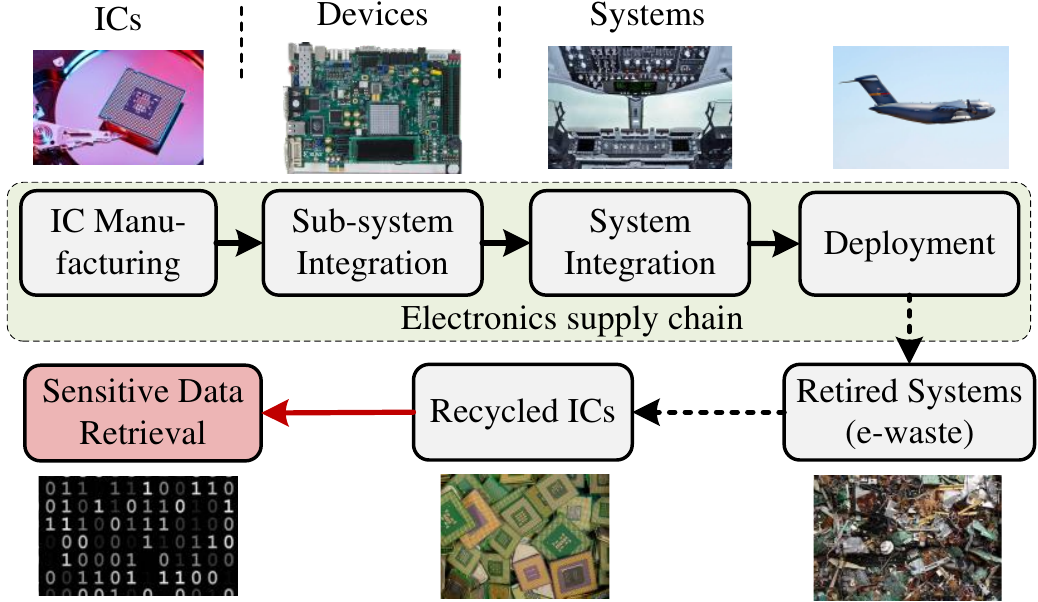}  \vspace{-15px}
\caption{\small The globalized electronics supply chain with threats.}
\label{fig:intro} \vspace{-10px}
\end{figure}

In parallel, recycling of ICs from electronic waste (e-waste) challenges researchers to accurately detect parts of inferior quality and reliability~\cite{guin2014counterfeitProc, guin2014counterfeitJETTA, tehranipoor2015counterfeit}. In the recycling process, electronic components are taken off from the printed circuit boards (PCBs), where most of these parts are still functional. Typically, the operation life of a chip is much larger than usage in the field~\cite{guin2014counterfeitProc, tehranipoor2015counterfeit}. The threats of these parts being used in critical infrastructures have been widely studied and explored by the researchers as they may have many defects and anomalies, and fail to deliver the desired performance when needed~\cite{congress2011counterfeit, guin2014comprehensive, tehranipoor2015counterfeit, leblanc2016nasa}. Unfortunately, these used ICs can also hold sensitive information related to the IP (e.g., the firmware/software, encryption keys, etc.) which can be exposed to an adversary. Figure~\ref{fig:intro} describes the overall electronics supply chain from IC manufacturing, sub-system/device integration, system integration, and deployment. Once the electronics are retired and discarded, the adversary recycles the working parts and can have the potential to recover previously-used data/programs from static random-access memories (SRAMs). \textit{To the best of our knowledge, this research is the first attempt to investigate the data retrieval from aged SRAM chips under normal operations.} Our objective is to show the hidden danger of discarding electronics when they are retired from our critical infrastructure. 

Data sanitization for non-volatile memories (NVMs) is well documented as these memories contain the firmware/software/ data for a typical computing system~\cite{hasan2020data}. NIST SP 800-88 Rev. 1 Guidelines for Media Sanitization~\cite{regenscheid2015nist} provides different permanent data erasure processes for NVMs, \eg, hard disk drive (HDD), solid-state drives (SSD), CDs, flash memory, memory sticks, etc. Unfortunately, there are no standard aging-related sanitization processes for volatile memories. SRAMs, a commonly used volatile memory, have a wide range of applications and are prevalent in electronic devices. They can be used as cache memories for processors or block RAMs for FPGAs, which hold sensitive data or programs once they are powered on. As it is a volatile memory, the most common perception is that it lost all its data after power off~\cite{NSA2014storage}. NSA/CSS Storage Device Sanitization manual mentions that the SRAM sanitization is instantaneous after the removal of power supply~\cite{NSA2014storage}.

In this paper, we present a novel approach for data retrieval from SRAM arrays that are previously used with fixed data, such as IPs or encryption keys. This is due to the effect of negative-bias temperature instability (NBTI) on PMOS transistors of the SRAM cell that ages towards the opposite value of the stored bit. Our partial data retrieval approach can recover the data by comparing the final power-up (FPU) state of a recycled SRAM chip with its initial power-up (IPU) state. Our threat model assumes that an adversary, an untrusted IC/device manufacturer, keeps the record of IPU states. Our improved partial data retrieval combines the recovered data from multiple chips and performs majority voting for better reconstruction of the sensitive data. The experimental result confirms our analysis and can recover the information previously stored in SRAM in as little as 4 hours of accelerated aging with 85$^{\circ}$C. \textit{We believe our work offers a novel perspective on the non-volatile property of the aged volatile SRAMs and sheds light on the importance of standardizing volatile memory sanitization protocols that can truly protect sensitive information from being unintentionally leaked to potential adversaries.} The contributions of this paper are summarized as follows: %very good

\begin{itemize}
    \item \textit{Novel data recovery hypothesis:} We propose a novel approach to determine the IP data that has been stored in the SRAM arrays. The threshold voltage ($V_{th}$) for one PMOS, not both, inside an SRAM cell, increases when 1-bit information is stored. With this change, the $V_{th}$ difference between the two PMOS transistors, initially present due to the process variation, either gets smaller or larger as the cell ages. Our analysis shows that aging pushes each SRAM cell toward a power-up state complementary to the previously stored value.
    \item \textit{Partial data retrieval:} As aging helps SRAM cells bias toward the opposite value of the stored data, we propose a partial data retrieval method that compares the SRAM's IPU state with the FPU state to determine the previously written SRAM content. 
    \item \textit{Improved data retrieval using multiple SRAM chips:} In addition to the partial data retrieval using one SRAM chip only, we propose the improved approach by combining the partially recovered data from multiple chips so that more data can be recovered. We demonstrate the effectiveness of this approach using six commercial-off-the-self SRAM chips aged with an image.
\end{itemize}

The rest of the paper is organized as follows. We briefly introduce the prior work in SRAM aging in Section~\ref{sec:background}. Our proposed approach for data retention analysis on the original data stored in SRAM chips is presented in Section~\ref{sec:proposed}. The experimental result and analysis for the proposed approach are described in Section~\ref{sec:result}. Finally, we conclude the paper in Section \ref{sec:conclusion}.

\section{Background}\label{sec:background}
The effects of process variation and aging on the power-up states of SRAM arrays have been extensively studied. Over the years, researchers have proposed physically unclonable functions (PUFs)~\cite{guajardo2007fpga, holcomb2009power, wang2022systematic} and true random number generators (TRNGs)~\cite{holcomb2009power, van2012efficient, wang2020aging} using SRAMs for creating unclonable device identifiers (IDs), and generating on-chip encryption keys. Due to the process variation, it becomes feasible to extract unique device fingerprints and create random numbers from an SRAM array. It is also well understood that the  aging degradation resulting from bias temperature instability (BTI)~\cite{schroder2003negative, reddy2005impact, wang2009impact} and  hot carrier injection (HCI)~\cite{chen1985reliability, mahapatra2006generation} influences the power-up state of an SRAM array, and affects the reliability of SRAM-based PUFs and TRNGs. Besides, it has been demonstrated that recycled ICs can be detected using the same power-up states that create SRAM PUFs and TRNGs~\cite{guin2019detecting}. This section presents a brief description of SRAM PUFs and TRNGs, and their usage in detecting recycled ICs.

\begin{figure}[t!]
\centering 
\includegraphics[width=\linewidth]{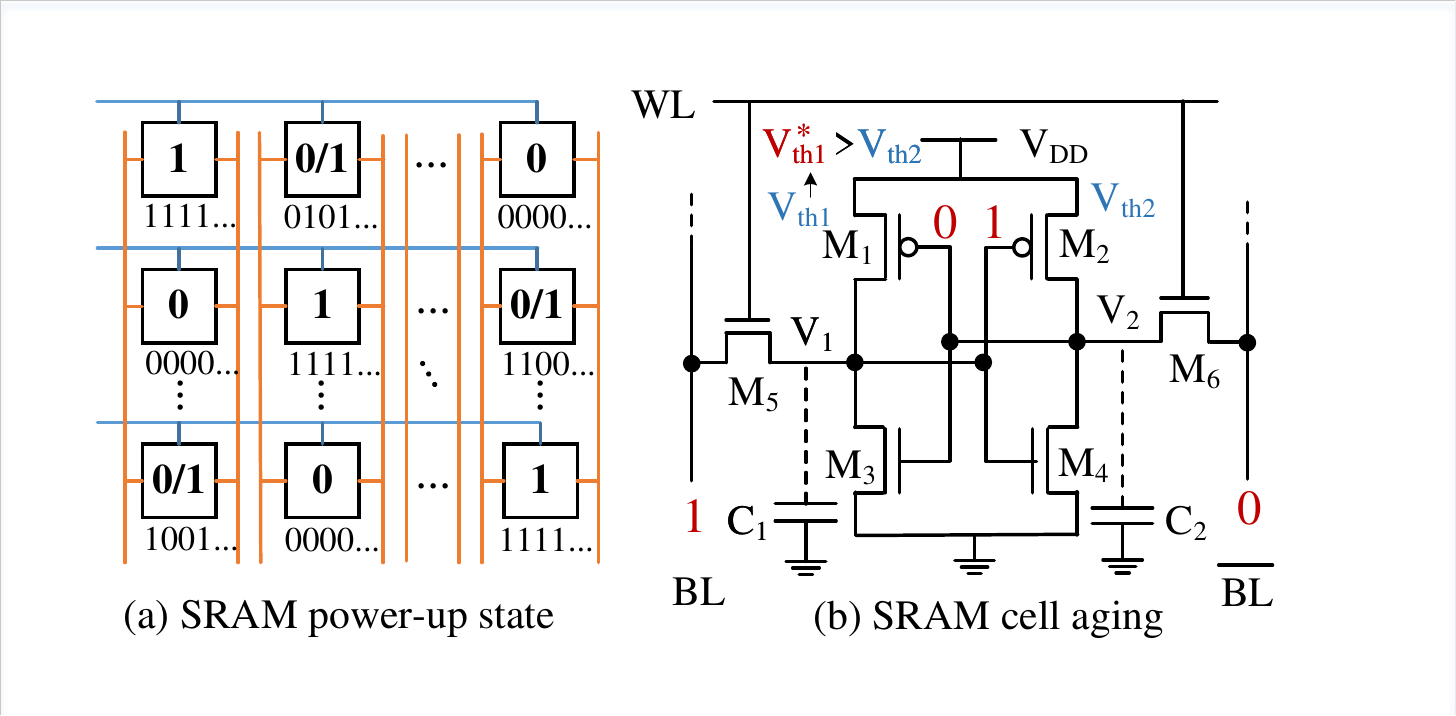} \vspace{-15px}
\caption{\small An abstract view of an SRAM array and an SRAM cell. (a) SRAM power-up state for creating unique ID~\cite{wang2022systematic, wang2020aging} and recycled IC detection~\cite{guin2019detecting}, (b) Effect of aging on a SRAM cell.}
\label{fig:SRAM_model} \vspace{-10px}
\end{figure}

\subsection{SRAM PUFs and TRNGs}\label{subsec:puf-trng}
Researchers have developed various PUF and TRNG structures utilizing SRAM arrays~\cite{guajardo2007fpga, holcomb2009power, van2012efficient, aysu2015end, mathew2016mu, wang2022systematic, wang2020aging, maes2013physically, xiao2014bit}. An SRAM cell, which typically consists of two back-to-back inverters, as shown in Figure~\ref{fig:SRAM_model}(b), is designed to be perfectly symmetrical to achieve the maximum static noise margin. However, process variation makes the cell biased toward one inverter, which results in a constant value in multiple power-ups. In addition, noise can play a role in flipping the power-up state. As a result, three types of power-up values can be observed, and they are: stable at logic 1 ($S_1$), \eg cells with `111...' power-up states in Figure~\ref{fig:SRAM_model}(a); stable at logic 0 ($S_0$), \eg cells with `000...' in Figure~\ref{fig:SRAM_model}(a); and unstable cells. A PUF can be realized when we extract the stable bits~\cite{holcomb2009power, van2012efficient, herder2014physical, wang2022systematic, wang2020aging}. This can create a unique/unclonable ID, as these stable cells reside in different locations and have unique internal biases for different chips due to random process variation. It is often required to incorporate error correction mechanisms~\cite{maes2013physically, xiao2014bit, schrijen2012comparative, herder2014physical} so that the ID remains stable as the chip ages in normal operation as well as under different environmental variations. 
Conversely, TRNG uses the unstable bits to generate random numbers~\cite{holcomb2009power, van2012efficient}, where the thermal and power supply noise has more effect on cells than the internal biases from manufacturing. 

\subsection{SRAM for detecting recycled ICs}\label{sec:sram-recycledic}

Figure~\ref{fig:SRAM_model}(b) shows an six-transistor SRAM cell, where the four transistors, $M_1, M_2, M_3$ and $M_4$, form a bistable latch to store 1-bit of data and $BL$/$\overline{BL}$ provide the access to the latch through $M_5$ and $M_6$ transistors. Guin et al. recently showed that the power-up state of an SRAM cell depends on the threshold voltages ($V_{th}$) of the MOS transistors, and how the power-up state depends on the content that an SRAM cell is aged with~\cite{guin2019detecting}. A new chip can be identified by observing an equal distribution of $1s$ and $0s$, as manufacturing process variation that shifts the $V_{th}$ of different transistors is Gaussian in nature. However, this equilibrium of  $1s$ or $0s$ gets skewed, and the percentage of $1s$ increases when an SRAM is aged with a content of more $0s$, commonly observed in programs and data. This can be explained using Figure~\ref{fig:SRAM_model}(b), due to the impact of negative bias temperature instability (NBTI) in traditional bulk technologies \cite{schroder2003negative, reddy2005impact}, PMOS transistor's $V_{th}$ increases over time. If logic 1 is stored in an SRAM cell, the $V_{th}$ of transistor $M_1$ will increase to $V^*_{th1} (>V_{th1})$ as it experiences NBTI-stress. At this point, the SRAM cell is now biased towards $M_2$ and will power up with logical 0 as $V^*_{th1}>V_{th2}$ . In summary, an SRAM cell will power up with logic 1 if we age the cell with logic 0 (and vice versa). One can find a detailed analysis of the effect of aging in the SRAM power-up state in~\cite{guin2019detecting}.

\section{Proposed Data Retrieval Approach from Used \\ SRAMs}\label{sec:proposed}

This section presents a novel hypothesis to reverse engineer the original data stored in the SRAM chip. As the aging increases the PMOS threshold voltage, the power-up state of each SRAM cell shifts toward the opposite of its stored information. An adversary can partially retrieve the previously stored sensitive data by simply reading the power-up state of SRAMs from the recycled SRAM arrays. We propose two approaches to partially recover the proprietary information that was once stored in SRAMs. The first approach targets the data retrieval for single SRAM chips, while the second approach merges the findings of multiple chips to improve the overall sensitive information retrieval. We begin this section by introducing the threat model.

\subsection{Threat model}

The threat model assumes that an adversary has long-term malicious objectives~\cite{whitehouse2021, ezell2021moore} to retrieve sensitive information from critical infrastructures. The threat model describes the capability of an adversary, and available resources to retrieve sensitive information from a recycled chip with on-chip SRAMs or commercial-off-the-shelf SRAM chips.  

\begin{itemize}
    \item \textit{Untrusted manufacturing:} The globalization of the electronics supply chain leads to offshore production of ICs and electronic devices. We consider electronics manufacturers as untrusted as these chips and devices are manufactured and assembled with limited trust and oversight. 
    \item \textit{Recycling of e-waste:} As e-waste is sent to offshore locations, the adversary can take hold of the used devices and once again read the SRAM content as these chips are still functional.
    
    \item  \textit{Access to initial power-up state information of SRAM arrays:} The adversary has the initial power-up (IPU) state for the SRAM arrays if they are meant for use in our critical system. As the manufacturing assembly, test and packaging, and sub-system/device integration are performed at offshore facilities, the initial power-up data of a new chip can easily be captured after its manufacturing.
    
    \item \textit{SRAM usage:} It is well understood that the secret information, e.g., the program code, resides in the same locations of an SRAM array during the normal operation. As a result, it is aged with the same content during its lifetime. 
    
    \item \textit{Data Sanitization:} Once the electronic device with SRAMs is retired from operation, it is simply discarded as SRAM is a volatile memory, and a typical sanitization is performed by powering down the chip~\cite{NSA2014storage}. 
\end{itemize}
%End of this subsection

\subsection{Hypothesis for SRAM Cell Data Recovery}\label{subsec:dataretrievalconcept} %\todoYadi{Add a few sentences}
\begin{figure}[t]
\centering 
\includegraphics[width=\linewidth]{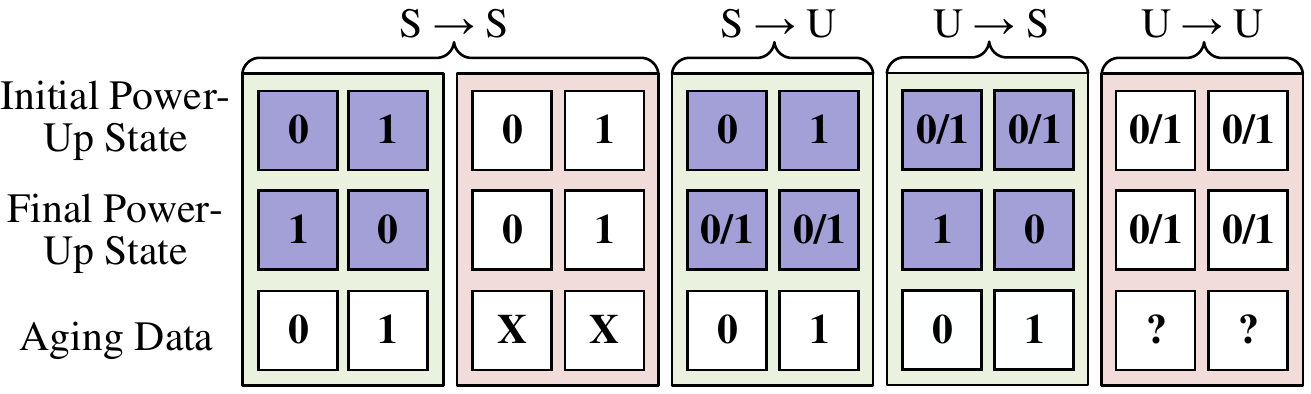} \vspace{-15px}
\caption{\small Aging data recovery from SRAM cells.} \vspace{-10px}
\label{fig:agingContent}
\end{figure}

As described in Section~\ref{sec:sram-recycledic}, the SRAM power-up state ages inversely to the binary data (or IP). As a result, it is possible to determine the actual SRAM content by observing the difference between the before- and after-aging power-up states. To identify the state accurately, SRAMs are generally powered on multiple times, as an SRAM array contains both stable ($S$) and unstable ($U$) cells. If a cell has consistently powered up with logic 0 (or 1), we call it a stable-0 ($S_0$) or stable-1 ($S_1$) cell. If the power-up content appears in both logic 0 and 1 in multiple power-ups, the cell is identified as an unstable cell. Four scenarios can happen for the stability change of each SRAM cell during aging. Figure~\ref{fig:agingContent} shows all four possible combinations of the IPU state and FPU state depending on the aging content (e.g., IP data). These four categories are explained as follows:
% \todoYadi{Describe the figure.}
\begin{itemize}
    \item \textit{Stable remains stable (S$\rightarrow$S):} An SRAM cell can be stable after aging either at the same or complementary initial logic values. If the FPU state (1 or 0) of a cell is opposite to its initial state (0 or 1), it represents that the aging has overcome the initial bias in a threshold voltage difference ($\Delta V_{th}$) of the PMOS transistors and has skewed towards the opposite direction. This only happens when the aging data is 0 for $S_0$ (becomes $S_1$) or 1 for $S_1$ (becomes $S_0$). However, if an SRAM cell remains at the same initial state, we cannot uniquely determine the aging content. For example, if both the IPU and FPU states remain 0, it is possible that either the cell has been aged with: ($i$) logic 0 but is not sufficient to offset the initial $\Delta V_{th}$ bias caused by process variation, or ($ii$) logic 1 that further increases the initial bias. Similar analysis can be done for logic 1 stable cells.
    
    \item \textit{Stable becomes unstable (S$\rightarrow$U):} A stable SRAM cell becomes unstable after usage in the field when aging balances the initial $\Delta V_{th}$ bias in the PMOS transistors. There is no ambiguity in determining the aging data as it helps shift the cell from stable to unstable. \textit{The aging data must be the initial state of the SRAM cell.}
    
    \item \textit{Unstable becomes stable (U$\rightarrow$S):} An unstable SRAM cell becomes stable after aging when it creates an imbalance at the final $\Delta V_{th}$ bias in the PMOS transistors. Like before, the aging contents can be uniquely determined. \textit{The aging data must be complementary to the final state of the SRAM cell.}
    
    \item \textit{Unstable remains unstable (U$\rightarrow$U):} If both the initial and final states of a cell remain unstable, aging may or may not have shifted the initial $\Delta V_{th}$ bias of the PMOS transistors. In order to determine whether aged data may be recovered from these cells, the distribution of the initial and final power-up states must be analyzed over multiple power cycles. If the distribution of 1 and 0 changes very little, then there is insufficient information to determine the aging data. However, if the distribution changes above a certain threshold, the data can be determined as the value which is closer to the original distribution. 
    
\end{itemize}

\subsection{Proposed Flow for SRAM Data Retrieval} \label{subsec:flow}
Since the majority of ICs and electronic devices are manufactured offshore, extra care must be taken to protect the sensitive information even though it has been erased permanently from all non-volatile memories once an electronic system retires. An impression of the stored data in a volatile memory, such as an SRAM, can be retrieved by an adversary. Figure~\ref{fig:adversary} shows the overall flow that describes data recovery from a used SRAM chip since its production and is presented using the following steps:

\begin{figure}[t!]
\centering 
\includegraphics[width=\linewidth]{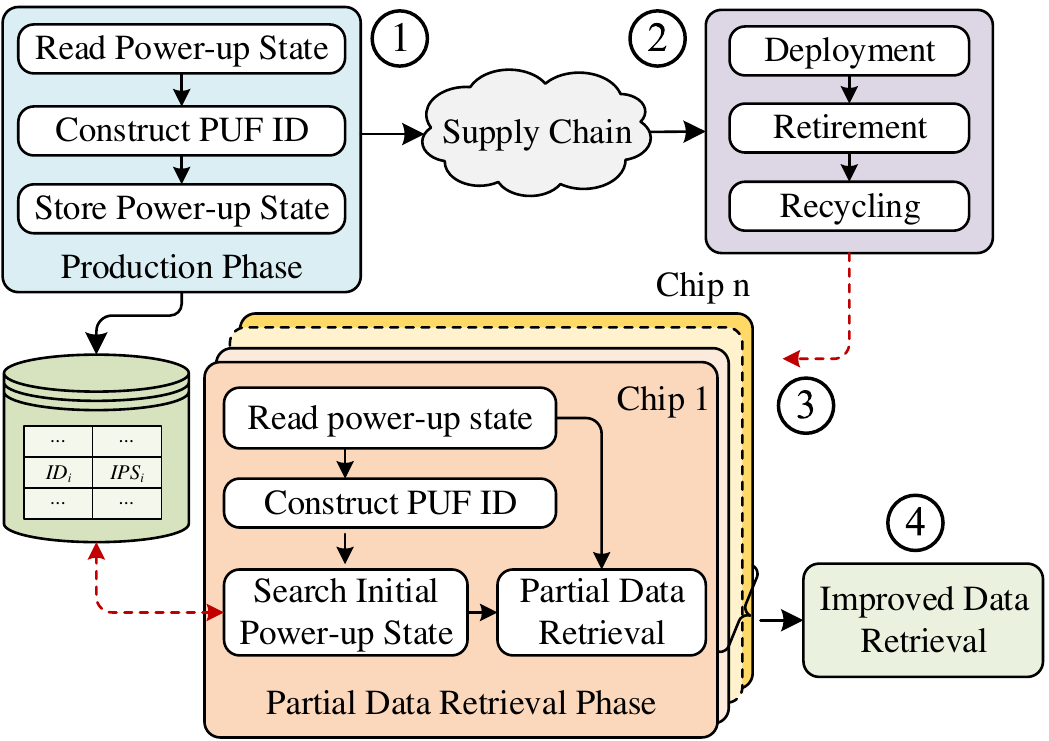} \vspace{-20px}
\caption{\small The proposed approach for recovering IP data.} % improved data {retrieval}
\label{fig:adversary} \vspace{-10px}
\end{figure}

\begin{itemize}
    \item \textit{Step 1 -- Preservation of initial power-up states after production:} To retrieve the aging data from an SRAM chip, an adversary needs to have access to the \textit{IPU} state. To uniquely identify the \textit{IPU} state for a particular chip, it needs to be attached with a unique ID. The ID can be constructed from the same \textit{IPU} state as described in Section~\ref{subsec:puf-trng}. An untrusted manufacturer of ICs/Devices can maintain a database of the $\{ID, IPU\}$ pairs. The step {\footnotesize \circled{1}} in Figure~\ref{fig:adversary} describes unique ID generation and storing it with \textit{IPU} in a database for future references. 
    
    \item \textit{Step 2 -- Recycling:} After the production phase, the electronics travel through the supply chain and are deployed in critical infrastructure. Note that the program data is, typically, loaded in the same locations of an SRAM chip. As a result, the chip gets aged with the same content. As these devices share the same function, they can be used in the same application, where these chips age with the same program data. Once the electronic devices are retired from operations, they are simply discarded rather than destroyed. As the e-waste is sent back to the same untrusted location, the same adversary can get hold of these devices, as shown in step {\footnotesize\circled{2}}.
    
    \item \textit{Step 3 --Partial data retrieval:} Once an adversary has access to the physical device, he/she first powers up the SRAM chip, reads the final power-up (FPU) states, and creates the unique device ID, as described in Section~\ref{subsec:puf-trng}. The IPU states can now be obtained by querying the existing database with this newly constructed ID. The adversary can compare both IPU and FPU states and partially reconstruct the data previously aged in the SRAM chip. The partial data retrieval phase is shown in {\footnotesize \circled{3}} of Figure~\ref{fig:adversary} and will be described in Section~\ref{subsec:partial} in details.
    
    \item \textit{Step 4 --Improved data retrieval:} The adversary can retrieve more aging data once he/she collects additional chips of the same type. The same {\footnotesize \circled{3}} is applied to all SRAM chips to collect partial data, and then use majority voting to construct complete aging data. The step {\footnotesize \circled{4}} in Figure~\ref{fig:adversary} shows the improved data retrieval, which will be described in Section~\ref{subsec:improved-partial} in details. 
\end{itemize}

\vspace{-7px}
\subsection{Partial Data Retrieval}\label{subsec:partial}\vspace{-3px}
As the adversary obtains both the IPU and FPU states for the SRAM chip of interest, he/she will apply the data recovery hypothesis described in Section~\ref{subsec:dataretrievalconcept} to each SRAM cell to retrieve its aging content. The SRAM data for IPU to FPU states change from stable 0 to stable 1 ($S_0\rightarrow S_1$)  or vice versa ($S_1 \rightarrow S_0$), stable to unstable ($S \rightarrow U$), or unstable to stable ($U \rightarrow S$) can be uniquely determined. Any unstable to unstable flip that leads to the power-up state shift above the threshold ($T_H$), the SRAM data is identified as well. If IPU and FPU states remain the same or unstable to unstable within the threshold limit, the data bit is indeterminate and cannot be recovered. %good

\begin{figure}[h!]
\centering \vspace{-10px}
\includegraphics[width=.9\linewidth]{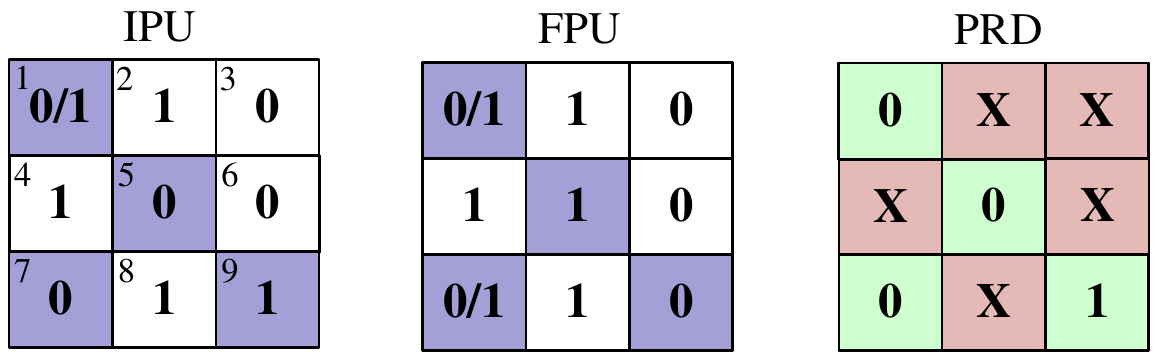}\vspace{-5px}
\caption{\small An example of partial data retrieval with an SRAM chip.}
\label{fig:partial-example}\vspace{-5px}
\end{figure}

Let us consider a simple example of partial data retrieval on a 3-by-3 SRAM cell, which is shown in Figure~\ref{fig:partial-example}. First, consider cell 5, which becomes $S_1$ from $S_0$ after aging. This becomes possible when the cell is aged with 0, and thus the recovered data is 0. Second, cell 1 is unstable (say, 20\% of 1s) initially and remains unstable after aging.  However, the cell now powers up 80\% of the time with 1s. The aging data of this cell must be logic 0 as the power-up state becomes more biased toward logic 1. Finally, the initial and final power-up content is stable logic 0 for cell 3. As a result, we cannot make any decision about its aging content. All these three analyses can be performed on all the other cells to compute the complete partially recovered data (PRD) from the IPU and FPU states.

\subsection{Improved Data Retrieval using Multiple SRAM Chips}\label{subsec:improved-partial}
Combining multiple PRDs can effectively increase the total recovered information from SRAM chips aged with the same content. First, practically half of the cells are aged with contents that help increase the initial threshold voltage bias $\Delta V_{th}$ of the PMOS transistors due to Gaussian random process variation~\cite{guin2019detecting}. It is practically impossible to determine the aging content due to the fact that stable cells become more stable. As a result, one cannot completely recover aging data from a single chip. If we combine multiple chips data, there will be a high probability that some of these cells will flip after aging, and we can recover their aging data. Second, our proposed approach uses majority voting on PRDs to reduce errors. It can be possible that the recovered content from one cell location from two/more chips may have conflicting values, which will create an error in data recovery. Considering multiple chips, we apply majority voting (a binary decision rule that selects the outcome based on majority) to remove the conflicts.

\begin{figure}[h!]
\centering  \vspace{-10px}
\includegraphics[width=\linewidth]{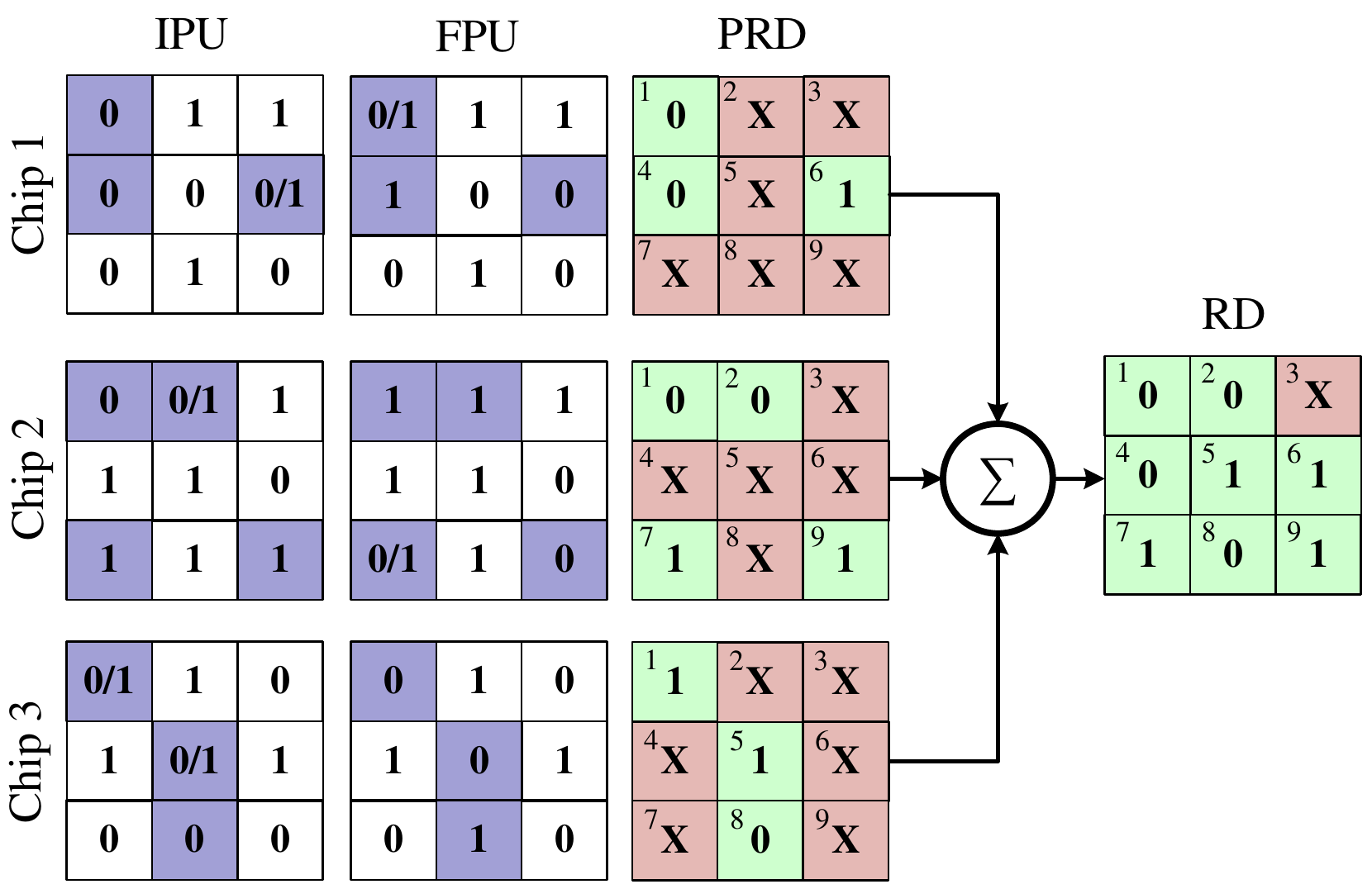}\vspace{-5px}
\caption{\small An example of the improved data retrieval using multiple SRAM chips.} \vspace{-5px}
\label{fig:improved-example}
\end{figure}

Figure~\ref{fig:improved-example} shows an example to explain our proposed scheme for improved data retrieval. First, cell 2 of chips 1, 2, and 3 contain `X', `0', and `X', respectively. This signifies that we can recover the cell 2 value from chip 2, where aging reinforces stability for chips 1 and 3. Second, for cell 1, we have conflicting recovered data. The data recovered from chips 1 and 2 is 0 while 1 for chip 3. The majority voting decides the recovered value of this cell as logic 0. The same two analyses can be performed throughout all SRAM cells to recover the aging data.

\setlength{\textfloatsep}{5pt}
\begin{algorithm}[!t]
\SetKwInOut{Input}{Input}\SetKwInOut{Output}{Output}
\SetKwProg{FPDR}{$\mathtt{part}$-$\mathtt{data}$-$\mathtt{rec}$}{ is}{end}
\SetKwProg{CHBA}{$\mathtt{hyp}$-$\mathtt{bit}$-$\mathtt{array}$}{ is}{end}
% \SetKwProg{AD}{$\mathtt{acc}$-$\mathtt{data}$}{ is}{end}
\SetAlgoLined

\Input{~Sets of IPU states ($\{IPU\}_M$), and FPU states\\ ($\{FPU\}_M$) for $N$ chips}
\Output{~Recovered data ($RD$)} 
\vspace{-5px}
\nonl \rule{.45\textwidth}{0.4pt}

\SetAlgoLined
\textbf{function} \FPDR{($\{IPU\}_M$, $\{FPU\}_M$)}{
        {\nonl \color{blue}/* ------ Partial Data Retrieval (Section~\ref{subsec:partial}) ------ */}\\ 

        $IS \leftarrow \varnothing$; $FS \leftarrow \varnothing$ \;
        $IS\leftarrow \mathtt{element}$-$\mathtt{wise}$-$\mathtt{addition}(\{IPU\}_{M})$ \;
        $FS \leftarrow \mathtt{element}$-$\mathtt{wise}$-$\mathtt{addition}(\{FPU\}_{M})$ \;
        $D\leftarrow\mathtt{diff}(IS, FS)$ \;
     	$H \leftarrow \mathtt{hyp}$-$\mathtt{bit}$-$\mathtt{array}(D)$ \;
     	return $H$ \;
}

\textbf{function} \CHBA{($D$)}{
    \ForEach{(element $[i,j] \in |D|$)}{
        \uIf{$D[i,j] > T_H$}{ 
            $H[i,j]\leftarrow 1$ ; {\color{blue}/* -- logic 1 -- */}\;}
        \uElseIf {$D[i,j] < -T_H$}{
            $H[i,j]\leftarrow -1$ ; {\color{blue}/* -- logic 0 -- */}\;
        }
        \Else{
            $H[i,j]\leftarrow 0$ ; {\color{blue}/* - insufficient information - */}\;
        }
            return $H$ \;
        
    }
}
{\nonl \color{blue}/* ------ Improved Data Retrieval (Section~\ref{subsec:improved-partial}) --------*/} \\

\For {$k\gets0$ \KwTo $N$ }{
    $H^k$=$\mathtt{part}$-$\mathtt{data}$-$\mathtt{rec}(\{IPU\}^k_M, \{FPU\}^k_M)$\;
}
{\nonl \color{blue}/* --------------- Majority Voting for $\pm$1 -------------- */}\\%expand

$RD\leftarrow \varnothing$ \; %I am ok..No
$PRD \leftarrow \mathtt{element}$-$\mathtt{wise}$-$\mathtt{addition}(\{H\}^{N})$ \;

\ForEach{(element $[i,j] \in |PRD|$)}{
        \uIf {$(PRD[i,j]) \geq +1$}{ 
        $RD[i,j] \leftarrow 1$ \; 
        }
        \uElseIf {$(PRD[i,j]) \leq -1$}{
            $RD[i,j] \leftarrow 0$ \;
        }
        \Else{
            $RD[i,j] \leftarrow X$ \;
        }
}
return $RD$ \;
\caption{The proposed SRAM data retrieval.} \label{alg:reconstruction} 
\end{algorithm} 

\subsection{Algorithm for Partial and Improved Data Retrieval}
Algorithm~\ref{alg:reconstruction} shows the overall improved data retrieval process (described in Section~\ref{subsec:flow}) where Lines 1-20 describe the partial one (shown in Section~\ref{subsec:partial}). For a recycled chip, multiple power-ups for both IPU and FPU states are recorded to capture the stable and unstable cells. We denote the number of power-ups as $M$, the collection of all $M$ IPUs and FPUs of the same chip as $\{IPU\}_M$, and $\{IPU\}_M$, respectively. The partial data retrieval ($\mathtt{part}$-$\mathtt{data}$-$\mathtt{rec}$) function takes $\{IPU\}_M$ and $\{FPU\}_M$ as inputs, Line 1. The initial and final state ($IS$ and $FS$) of an SRAM chip can be computed as the element-wise addition of power-up states for all $M$ measurements, Lines 3-4. As described in the data retrieval concept in Section~\ref{subsec:dataretrievalconcept}, a chip's aging bias is the complementary imprint of the IP data written to SRAM cells. If a chip is aged with logic 0, the final power-up will lean towards more 1s, \eg a change of 5 to 9 from $IS$ to $FS$ under $M=10$, opposite to the programmed data bit. This inverse shift in SRAM cell bias can be identified by computing the difference between the accumulated initial and final states, $IS$ and $FS$, Line 5. The hypothesis bit array ($H$), or the partially reconstructed data from a single chip, can be computed with function $\mathtt{hyp}$-$\mathtt{bit}$-$\mathtt{array}$, Lines 9-20. It examines the difference between $IS$ and $FS$ of each bit based on a sample-size-dependent positive threshold $T_H$. If the difference is positive and greater than $T_H$, we know it has been aged with data opposite to the aging bias of 0, that is, a logic 1 bit, also saved in array H, Lines 11-12. If the difference is less than the threshold $-T_H$, a shift toward 1 for a cell's power-up indicates that the cell must have been aged with a logic 0 data, where we mark it with -1 in $H$, Lines 13-14. If the difference between $IS$ and $FS$ is relatively insignificant to sample size $M$, we consider aging data as indeterminate and record it with 0 in $H$, Lines 15-17. The constructed bit array with thresholding is returned once the ternary information for all SRAM cells has been generated, Line 18. 

Lines 21-35 describe the improved data retrieval process using multiple ($N$) SRAM chips. Each chip contributes to data retrieval from certain unique cells which remain stable or become more stable after aging in other chips. Thus quite a few indeterminate bits will be recovered when an additional chip is added. In regard to conflicting values between multiple chips, \eg one chip determines a cell with data 0 while another chip calls it data 1, the majority voting takes place and will help to resolve the conflict, Lines 24-34. Any cells with a net zero sum in majority voting are considered inconclusive, which we marked as $X$, Lines 31-33. The recovered SRAM data ($RD$) is returned once the program completes the accumulation of partially retrieved data from multiple SRAM chips, Line 35.

\section{Experimental Results}\label{sec:result}
\begin{figure}[t]
\centering 
\includegraphics[width=\linewidth]{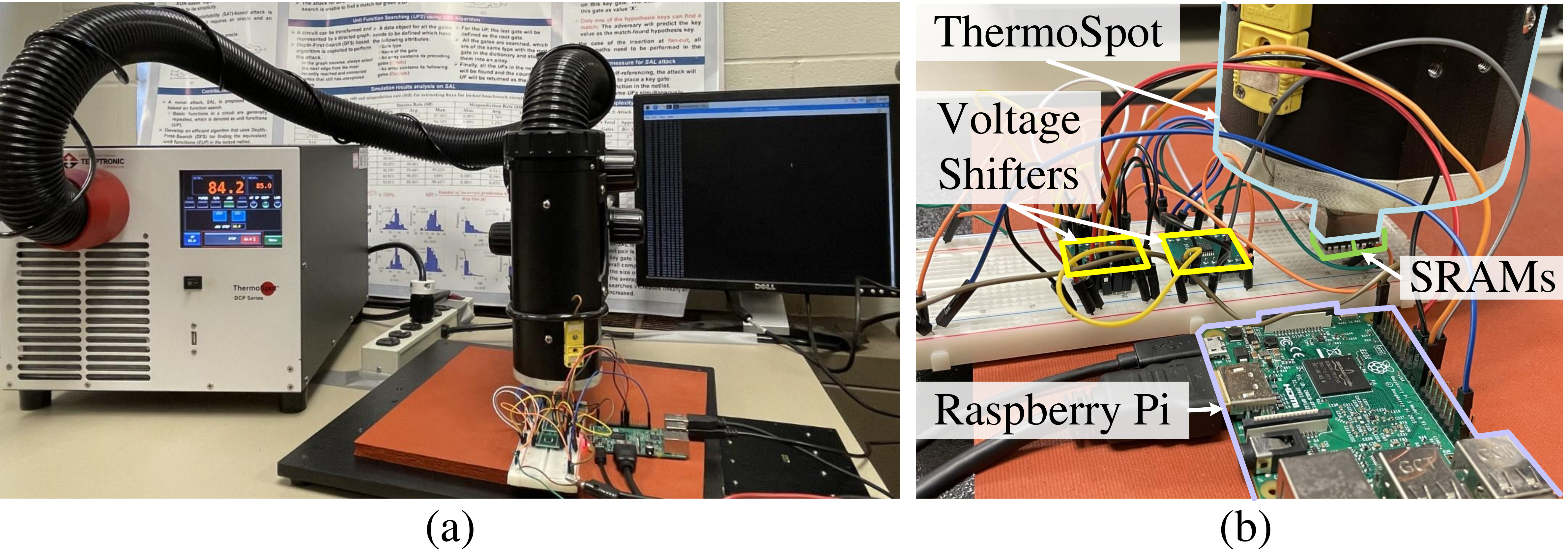} \vspace{-20px}
\caption{\small Experimental setup. (a) Accelerated aging set-up using ThermoSpot, (b) zoomed-in view of SRAM aging.} %\vspace{-10px}
\label{fig:setup}
\end{figure}

The sensitive information, stored in the SRAM during its operation, can be leaked to an adversary once he/she compares the initial and final power-up states. In this section, we demonstrate data retrieval from commercial-off-the-shelf SRAM chips. We age the chips by storing a black and white image. The reconstruction of SRAM data (Algorithm~\ref{alg:reconstruction}) is processed via in-house MATLAB script. Our proposed approaches can retrieve an imprint of the image in as little as 4 hours of accelerated aging.

Figure~\ref{fig:setup} shows the experimental setup for accelerated aging of SRAMs. The SRAM chips are aged with the contact probe of Temptronic ThermoSpot DCP-201 system~\cite{temptronic} at the constant 85$^{\circ}$C. Six commercial off-the-shelf Microchip 64 Kbit serial SRAMs~\cite{microchip8k} are selected to perform the experiment. All SRAMs are aged with the same black and white image. Multiple power-up states of an SRAM are recorded every 2 hours of aging. The data is collected from the Raspberry Pi through the SPI interface. Voltage shifters ensure voltage compatibility between the SRAM chips (1.65V) and Raspberry Pi (3.3V) during read/write of SRAM data.

\subsection{Partial data retrieval}

\begin{figure}[t]
\centering 
\includegraphics[width=\linewidth]{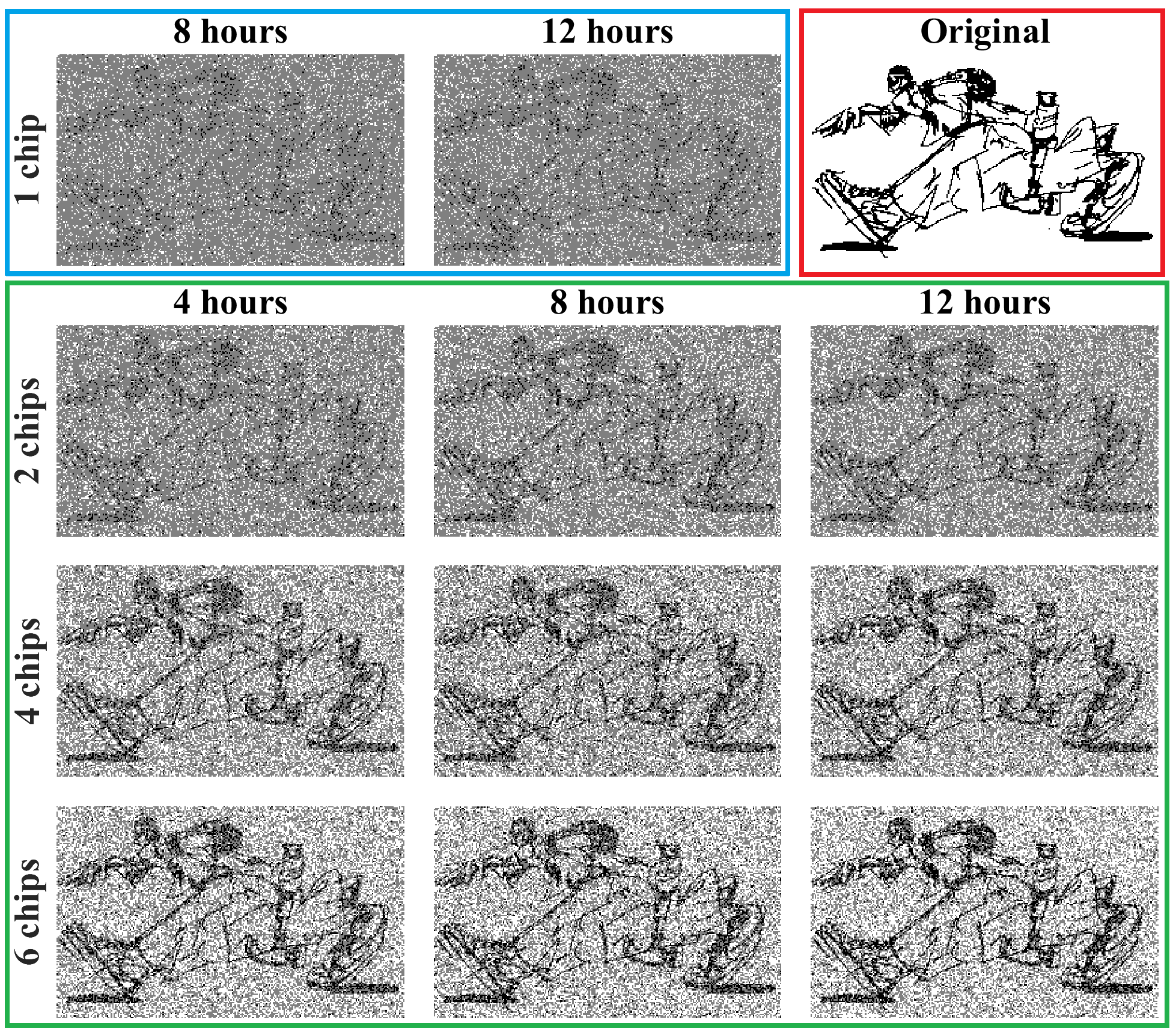}\vspace{-5px}
\caption{\small The partial data retrieval from one SRAM chip and improved reconstruction with multiple SRAM chips.}
\label{fig:result1}%\vspace{-10px}
\end{figure}

Our partial data approach focuses on the aging characteristics of each individual chip. The top row of Figure~\ref{fig:result1} shows the data retrieval from an individual SRAM chip aged for 8 and 12 hours. In the top rightmost, red box, the original image that is stored in the SRAM during aging is shown. The power-up states are recorded 10 times at 2-hour intervals. This allows for close monitoring of each chip’s aging development, and shows that as the amount of $V_{th}$ degradation from aging increases, the amount of data recovered increases as well. Even with only one chip, a clear impression of the original image can be observed after 12 hours of accelerated aging. As any SRAM chip contains a majority of stable cells (typically more than 80\%), aging will bias towards stability (stable becomes more stable) for nearly half of the cells, and we cannot extract aging content from those cells. This puts a hard cap on the amount of information that can be retrieved from just one chip.

\subsection{Improved data retrieval}

To improve the clarity of the image and reduce estimation noise, one needs to combine the retrieved data from multiple chips with a majority voting system. As a result, we can find the cells that are most prone to aging from each chip and combine them, lowering the amount of bits that are impossible (e.g., stable becomes more stable) to gain information from with each added chip. Additionally, considering the majority voting system, if the one bit on a chip creates a false output due to noise, another chip’s correct output will cancel it out. As the number of chips increases, the correct information will further outnumber the noisy information and remove it. Figure~\ref{fig:result1} shows the different outputs for 4, 8, and 12 hours of aging with varying numbers of chips, demonstrating that as the amount of chips increases, the clarity of the recovered data improves as well.

\section{Conclusion}\label{sec:conclusion}
In this paper, we have shown, for the first time, that the data written to SRAM chips is permanently stored and can be partially retrieved from the aged ones, although the original data has been removed after the chip is powered off. Our novel data retrieval concept provides the foundation for recovering data bits by analyzing the difference between the initial and final power-up states of an SRAM. We propose two approaches to retrieve the sensitive information previously stored in one or multiple SRAM chips. The experimental results demonstrate the successful reconstruction of a large portion of the original data even if the SRAM has been aged for a very short time. We believe our work is crucial for understanding and safeguarding the sensitive data that was once stored in critical applications, and it proves the need for a rigorous sanitization standard for SRAMs for both industrial and defense sectors. In our future work, we plan to develop a robust data retrieval scheme for SRAMs without having any knowledge of the initial power-up state and use only the final power-up states of multiple chips as self-referencing.

\section*{Acknowledgment} This work was supported by the National Science Foundation under Grant Number CNS-1755733. %Any opinions, findings, conclusions, or recommendations expressed in this material are those of the authors and do not necessarily reflect the views of the National Science Foundation.

% \pagebreak
% \bibliographystyle{ieeetr} %for ieee
% \bibliography{bib, guinbib}

\end{document}